\documentclass[conference]{IEEEtran}

\usepackage[symbol]{footmisc}
\usepackage{authblk}
\usepackage{diagbox}
\usepackage{threeparttable}
\usepackage{enumitem}
\usepackage{url}
\usepackage[vlined,ruled]{algorithm2e}
\usepackage{amsmath}
\usepackage{amssymb}
\usepackage{url}
\usepackage{graphicx}
\usepackage{subfig}
\usepackage{algorithmic}
\usepackage{tabu}
\usepackage{float} 
\usepackage{framed}
\usepackage{gensymb}
 \usepackage[noadjust]{cite}
\usepackage{array}
\usepackage{comment}
\usepackage{xspace}
\usepackage{afterpage}
\usepackage{listings}
\usepackage{xcolor}
\usepackage{caption} 

\definecolor{codegreen}{rgb}{0,0.6,0}
\definecolor{codepurple}{rgb}{0.58,0,0.82}

\lstdefinestyle{mystyle}{  
    commentstyle=\color{codegreen},
    keywordstyle=\color{magenta},
    numberstyle=\tiny\color{gray},
    stringstyle=\color{codepurple},
    basicstyle=\ttfamily\footnotesize,
    breakatwhitespace=false,         
    breaklines=true,                 
    captionpos=b,                    
    keepspaces=true,                 
    numbers=left,                    
    numbersep=5pt,                  
    showspaces=false,                
    showstringspaces=false,
    showtabs=false,                  
    tabsize=2
    }

\begin{document} 

\long\def\authornote#1{%
  \leavevmode\unskip\raisebox{-3.5pt}{\rlap{$\scriptstyle\diamond$}}%
  \marginpar{\raggedright\hbadness=10000
    \def\baselinestretch{0.8}\tiny
    \it #1\par}}
\newcommand{\justin}[1]{\authornote{Justin: #1}}
\newcommand{\zitao}[1]{\authornote{Zitao: #1}}
\newcommand{\karthik}[1]{\authornote{Karthik: #1}}
\newcommand{\NJ}[1]{\authornote{NJ: #1}}

\newcommand{\tfname}{TensorFlow\xspace}
\newcommand{\sysname}{TensorFI\xspace}

\newcommand\blfootnote[1]{%
  \begingroup
  \renewcommand\thefootnote{}\footnote{#1}%
  \addtocounter{footnote}{-1}%
  \endgroup
}

\title{\sysname: A Flexible Fault Injection Framework for TensorFlow Applications } 

\author[1]{Zitao Chen$^*$}
\author[1]{Niranjhana Narayanan$^*$}
\author[1]{Bo Fang}
\author[2]{Guanpeng Li}
\author[1]{Karthik Pattabiraman}
\author[3]{Nathan DeBardeleben}

\affil[1]{University of British Columbia}
\affil[2]{University of Illinois}
\affil[3]{Los Alamos National Laboratory}


 \maketitle

\def\thefootnote{*}\footnotetext{Equal contributions}
\blfootnote{A preliminary version of this work was published in a workshop~\cite{li2018tensorfi}.}

\begin{abstract}
As machine learning (ML) has seen increasing adoption in safety-critical domains (e.g., autonomous vehicles), the reliability of ML systems has also grown in importance. While prior studies have proposed techniques to enable efficient error-resilience techniques (e.g., selective instruction duplication), a fundamental requirement for realizing these techniques is a detailed understanding of the application's resilience.
In this work, we present \sysname, a high-level fault injection (FI) framework for TensorFlow-based applications. \sysname is able to inject both hardware and software faults in general TensorFlow programs. \sysname is a configurable FI tool that is flexible, easy to use, and portable. It can be integrated into existing TensorFlow programs to assess their resilience for different fault types (e.g., faults in particular operators). We use \sysname to evaluate the resilience of 12 ML programs, including DNNs used in the autonomous vehicle domain. 
Our tool is publicly available at \fbox{ \url{https://github.com/DependableSystemsLab/TensorFI}}.

\end{abstract} 

\section{Introduction}
In the past decade, Machine Learning (ML) has become ubiquitous in many applications.
ML is also being increasingly deployed in safety-critical applications such as Autonomous Vehicles (AVs)~\cite{banerjee2018hands}
 and aircraft control~\cite{julian2016policy}.
In these domains, it is critical to ensure the reliability of the ML algorithm and its implementation as faults can lead to loss of life and property.
Moreover, there are often safety standards in these domains that prescribe the allowable failure rate. 
For example, in the AV domain, the ISO 26262 standard mandates that the FIT rate (Failures in Time) of the system be no more than 10, 
i.e., at most 10 failures in a billion hours of operation~\cite{iso26262}, in order to achieve ASIL-D levels of certification. Therefore, there is a compelling need
to build efficient tools to (1) test and improve the reliability of ML systems, and (2) evaluate their failure rates in the presence of different fault types.

The traditional way to experimentally assess the reliability of a system is fault injection (FI). FI can be implemented at the hardware or software
level. Software-Implemented FI (also known as SWiFI) has lower costs, is more controllable, and easier for developers to deploy~\cite{hsueh1997fault}. 
Therefore, SWiFI has become the dominant method to assess a system's resilience to both hardware and software faults. 

There has been a plethora of SWiFI tools such as NFTape~\cite{stott2000nftape}, Xception~\cite{carreira1998xception}, GOOFI~\cite{aidemark2001goofi}, LFI~\cite{marinescu2009lfi},
LLFI~\cite{thomas2013llfi}, PINFI~\cite{wei2014quantifying}.
These tools operate at different levels of the system stack, from the assembly code level to the application's source code level.
In general, the higher the level of abstraction of the FI tool, the easier it is for developers to work with, and use the results from the FI
experiments~\cite{hsueh1997fault}. 

Due to the increase in popularity of ML applications, there have been many frameworks developed for writing them. One of the most
popular frameworks is \tfname~\cite{abadi2016tensorflow}, which was released by Google in 2017. Other examples are PyTorch~\cite{paszke2019pytorch} and Keras~\cite{keras}.
Unlike traditional applications, these frameworks allow the developer to ``compose'' their application as a sequence of {\em operations},
which are connected together in the form of  a graph. The connections represent the data-flow and control dependencies among the
operations. While the underlying implementation of these frameworks is in C++ or assembly code for performance reasons, the developer
writes their code using high-level languages (e.g., Python). 

In this paper, we introduce a SWiFI tool called \sysname
which injects faults into the data-flow graph used in \tfname applications. 
\sysname performs {\em interface-level FI}~\cite{kropp1998automated,lanzaro2014empirical}.
We focus on \tfname as it is {\em the most popular framework} used today for ML applications~\cite{tf-popularity}, though our technique is
not restricted to \tfname. \sysname can be used to inject both hardware and software faults in the outputs of \tfname operators, and study
the effects of the faults on the ML application. The main advantage of \sysname over traditional SWiFI frameworks is that it directly operates
on the \tfname operators and graph, and hence its results are readily accessible to developers. 

Building a FI tool for \tfname applications is challenging due to three reasons. First, because \tfname operators are implemented in C++ or 
assembly code and optimized for different platforms (i.e., different processors and operating systems), it is not practical to modify the
 implementation of these operators as doing so will hurt both portability and performance. 
 However, in order to inject faults at the level of \tfname operators and the graph, one needs to intercept the operators
at runtime to modify their execution results. Unfortunately, \tfname does not expose the operators once the graph has been constructed, and
most of the execution occurs ``behind the scenes'' in the low-level code. Therefore, it is not possible to intercept these operators. 
Secondly, the speed of execution of the \tfname graph should not be adversely affected when no faults are 
injected, as otherwise developers will avoid using the framework. 
Finally, there are many external libraries that are used by \tfname developers.
These often rely on the structure and semantics of the \tfname
graph, and hence these should not be modified.

\sysname addresses the above challenges by first duplicating the \tfname graph and creating a {\em FI graph} that parallels
the original one. The operators in the FI graph mirror the functionality of the original \tfname operators, except that they have
 the capability to inject faults based on the configuration parameters specified. These operators are implemented by us 
 in Python, thereby ensuring their portability. Moreover, the FI graph is only invoked during fault injection, and hence the 
 performance of the original \tfname graph is not affected (when faults are not injected). Finally, because we do not modify the
 \tfname graph other than to add the FI graph, 
 external libraries that depend on the graph's structure and semantics can continue to work. 
 
 We make the following contributions in this paper. 
 \begin{itemize}[leftmargin=*]
 \item Propose a generic FI technique to inject faults at the level of the \tfname graph, without hurting portability and performance,
 \item Implement the FI technique in \sysname, a flexible tool, which allows easy configuration of FI parameters.
 \item Evaluate \sysname on $12$ ML applications in \tfname, including deep neural network (DNN) applications used in AVs,  
  across a wide range of FI configurations (e.g., fault types, error rates).  We find that there
  are significant differences due to both individual ML applications, as well as due to different configurations.
  We also evaluate the performance of \sysname.
 \end{itemize}
 

\label{sec:introduction}
\section{Background and Fault Model}
\label{sec:background}
\label{sec:background}
We start by explaining the general structure of ML applications, followed by related work in the area of ML reliability. We then introduce the fault model we assume in this paper. 

\subsection{Machine Learning Applications}
\label{sec:background-deep-learning}
An ML model takes an input that contains specific features to make a prediction. Prediction tasks can be divided into classification and regression. The former is used to classify the input into categorical outputs (e.g., image classification). The latter is used to predict dependent variable values based on the input. ML models can be either supervised or unsupervised. In the supervised setting, the training samples are assigned with known labels (e.g., linear regression, neural network), while in an unsupervised setting there are no known labels for the training data (e.g., k-means, kernel density estimation). 

An ML model typically goes through two phases: 1) training phase where the model is trained to learn a particular task; 2) inference phase where the model is used for making predictions on test data. The parameters of the ML model are learned from the training data, and the trained model will be evaluated on the test data, which represents the unseen data and will not be used in the training phase. 

\subsection{Related Work}
\label{sec:related-work}

Several studies have attempted to evaluate the error resilience of ML applications through fault injections~\cite{alippi1995sensitivity,bettola1998high}. However, such FI techniques are limited to the specific application being studied, unlike \sysname that is able to perform FI on generic ML applications.

More recent studies investigate the resilience of deep neural networks (DNN) by building fault injectors~\cite{li2017understanding,reagen2018ares,chen2019,sabbagh2019evaluating}. Li et al. build a fault injector by using the tiny-CNN framework~\cite{li2017understanding}. Reagen et al design a generic framework for quantifying the error resilience of ML applications~\cite{reagen2018ares}. Sabbagh et. al develop a framework to study the fault resilience of compressed DNNs~\cite{sabbagh2019evaluating}. However, their study focuses only on DNNs and hardware faults, and hence is not applicable to other ML algorithms. Chen et al. introduce a technique to efficiently prune the hardware FI space by analyzing the underlying property of ML models~\cite{chen2019}. 
In contrast to the above studies, \sysname targets a broad range of ML applications, and can be used to inject both software and hardware faults. 

\subsection{\tfname}

\tfname is an open-source framework for modeling large data-flow graphs and widely used for building ML programs. 
\tfname allows programmers to represent the program in the form of a \tfname graph (see below). 
\tfname is flexible and can be better optimized as it exposes the underlying graph to the developer. Thus, \tfname is considered a low level framework. 
Many high level frameworks like Keras use \tfname as their backend for implementation. 

%


To use \tfname, programmers use the built-in operators to construct the data-flow graph of the ML algorithm during the {\em training phase}. 
Once the graph is built, 
it is not allowed to be modified. 
During the {\em inference phase}, data is fed into the graph through the use of placeholder operators, and the outputs of the graph correspond to the outputs of the ML algorithm. In this phase, the graph is typically executed directly in the optimized form on the target platform using custom libraries.

\tfname also provides a convenient Python language interface for programmers to construct and manipulate the data-flow graphs. Though other languages are also supported, the dominant use of \tfname is through its Python interface. Note however that the majority of the ML operators and algorithms are implemented as C/C++ code, and have optimized versions for different platforms. The Python interface simply provides a wrapper around these C/C++ implementations.

\subsection{Fault Model}
\label{sec:fault-model}

In this work, we consider two types of faults, hardware faults and software faults that occur during the execution of the \tfname program. As \sysname operates at the level of \tfname operators, we abstract the faults to the operators' interfaces. Thus, we assume that a hardware or software fault that arises within the \tfname operators,
ends up corrupting (only) the outputs of the operators. We do not make assumptions on the nature of the output's corruption. 
For example, we consider that the output corruption could be manifested as a random value replacement (e.g., mutation testing~\cite{ma2018deepmutation}) or a single bit-flip~\cite{li2017understanding,reagen2018ares,chen2019,sabbagh2019evaluating}. We also assume that the faults do not modify the structure of the \tfname graph,
 and that the inputs provided into the program are correct, because such faults are extraneous to \tfname and are outside our scope. 
 Finally, we assume that the faults occur neither in the ML algorithms, nor in the parameters of the ML model. 
 This assumption is needed for us to compare the output of the FI runs with the golden runs, in order to determine if a Silent Data Corruption (SDC) has occurred. 

We only consider faults during the {\em inference} phase of the ML program. 
Because training is usually a one-time process and the results of the 
trained model can be checked. Inference, however, is executed repeatedly with different inputs, and is hence likely to experience faults. 
This fault model is in line with other work in this area~\cite{li2017understanding,reagen2018ares,chen2019,sabbagh2019evaluating}. 

\section{Methodology}
We start this section by articulating the design constraints of \sysname. 
We then present the design of \sysname, and an example of its operation. Finally, we present its implementation and explain how to configure it.

\subsection{Design Constraints}

We follow 3 constraints in the design of \sysname.
\begin{itemize}[leftmargin=*]
\item {\bf Ease of Use and Compatibility}: The injector should be easy-to-use and require minimal modifications to the application code.  We also need to ensure compatibility with third-party libraries that may construct the \tfname graph using custom APIs.
\item {\bf Portability}: Because \tfname may be pre-installed on the system, and each individual system may have its own installation of \tfname, we should not assume the programmer is able to make any modifications to \tfname or its libraries.
\item {\bf Minimal Interference}: First, the injection process should not interfere with the normal execution of the \tfname graph when no faults are injected. Further, it should not make the main graph incapable of being executed on GPUs or parallelized due to the modifications it makes. Finally, the fault injection process should be reasonably fast, when faults are injected.
\end{itemize}

We also make two assumptions in \sysname.
First, we assume that faults occur only during the execution of the \tfname operators, and that the faults are transient in nature. In other words, if we reexecute the same operator, the fault will not reappear. This is because studies have shown that the kinds of faults that are prevalent in mature software are often transient faults~\cite{iyer1993experimental}. 
Second, we assume that the effect of a fault propagates to the outputs of the \tfname operators only, and not to any other state. In other words, there is no error propagation to the permanent state, which is not visible at \tfname graph level. Again, this is due to the structure of \tfname graphs, and our fault model (Section~\ref{sec:background}).

\subsection{Design of \sysname}

To satisfy the design constraints outlined earlier, \sysname operates directly on \tfname graphs. The main idea is to create a replica of the original \tfname graph but with new operators. The new operators are capable of injecting faults during the execution of the operators and can be controlled by an external configuration file. Further, when no faults are being injected, the operators emulate the behavior of the original \tfname operators they replace.
Because \tfname does not allow the dataflow graph to be modified once it is constructed, we need to create a copy of the entire graph, and not just the operators we aim to inject faults into. The new graph mirrors the original one, and takes the same inputs as it. However, it does not directly modify any of the nodes or edges of the original graph and hence does not affect its operation. At runtime, a decision is made as to whether to invoke the original \tfname graph or the duplicated one for each invocation of the ML algorithm. Once the graph is chosen, it is executed to completion at runtime. 


\sysname works in two phases. The first phase instruments the graph, and creates a duplicate of each node for fault injection purposes. The second phase executes the graph to inject faults at runtime, and returns the corresponding output. Note that the first phase is performed only once for the entire graph, while the second phase is performed each time the graph is  executed (and faults are injected).
We explain below how this satisfies the design constraints. 

\begin{itemize}[leftmargin=*]
\item {\bf Ease of Use and Compatibility}: To use \sysname, the programmer changec a single line in the Python code. Everything else is automatic, namely the graph copying and duplication. Because we duplicate the \tfname graph, our method is compatible with external libraries.

\item {\bf Portability}: We do not make any modifications to the \tfname code or the internal C++ implementation of the \tfname operators, which are platform specific. Therefore our implementation is portable.
We do however require the interfaces of the \tfname operators are consistent across \tfname versions. 

\item {\bf Minimal Interference}: \sysname does not interfere with the operation of the main \tfname graph. Further, the original \tfname operators are not modified in any way, and hence they can be optimized or parallelized for specific platforms. The only overhead introduced by \sysname is the check at runtime on whether to call the original graph or the duplicated graph, but this incurs minimal overhead. 
\end{itemize}

\subsection{Example of \sysname's Operation}

\begin{figure}
\centering \includegraphics[height=1.0in, width=0.4\textwidth]{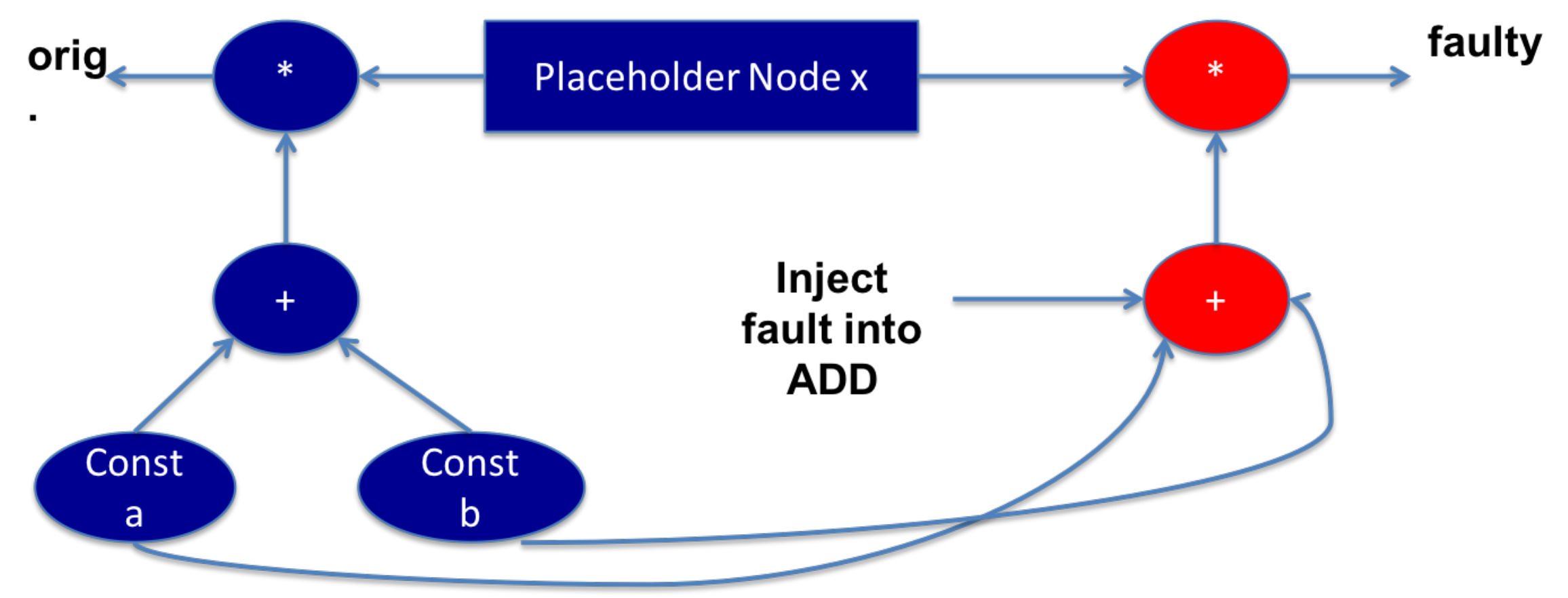}
\caption{Example of \tfname graph and how \sysname modifies it. The nodes in blue represent the original nodes in the graph, while the nodes in red are those added by \sysname for fault injection purposes.}
\label{fig:TensorFI-example}
\end{figure}

We consider an example of \sysname's operation on a small \tfname program. Because our goal is to illustrate the operation of \sysname, we consider a simple computation rather than an ML algorithm. The example is shown in Figure~\ref{fig:TensorFI-example}. The nodes in blue represent the original \tfname graph, while those in red represent the duplicated nodes created by \sysname. 

In the original \tfname graph, there are two operators, an ADD operator which adds two constant node ``a" and ``b", and a MUL operator, which multiplies the resulting value with that from a place-holder node. A place-holder node is used to feed data from an external source such as a file into a \tfname graph, and as such represents an input to the system. A  constant node represents a constant value. \sysname duplicates both the ADD and MUL operators in parallel to the main \tfname graph, and feeds them with the values of the constant nodes as well as the place-holder node. Note however that there is no flow of values back from the duplicated graph to the original graph, and hence the fault injection nodes do not interfere with the original computation performed by the graph. The outputs {\em orig} and {\em faulty} represent the original and fault-injected values respectively. The graph is created before the fault injection process is launched after the training phase. 

At runtime, a dynamic decision is made as to whether we want to compute the {\em orig} output or the {\em faulty} output. If the {\em orig} output is demanded, then the graph nodes corresponding to the original \tfname graph are executed. Otherwise, the nodes inserted by \sysname are executed and these emulate the behavior of the original nodes, except that they inject faults. For example, assume that we want to inject a fault into the ADD operator. Every other node inserted by \sysname would behave exactly like the original nodes in the \tfname graph, with the exception of the ADD operator which would inject faults as per the configuration (Section~\ref{sec:configurations}). 

\subsection{Implementation}
\label{sec:implementation}

\sysname supports the following features:
\begin{itemize}
\item Comparing each FI result with the golden run
\item Launching multiple FI runs in parallel (multi-threading)
\item Support for visualizing the modified \tfname graphs 
\item Ability to specify fault type etc. in a configuration file
\item Automated logging of fault injection runs
\item Support for statistics collection and analysis
\end{itemize}

Our implementation consists of about 2500 lines of heavily commented Python code, and is split into 5 modules. We have made \sysname publicly available under a MIT license on Github (\fbox{ \url{https://github.com/DependableSystemsLab/TensorFI}}),
along with extensive documentation. 



\label{sec:methodology}
\subsection{Configurations}
\label{sec:configurations}

\begin{figure}
    \centering
    \includegraphics[height=1.0in, width=0.3\textwidth]{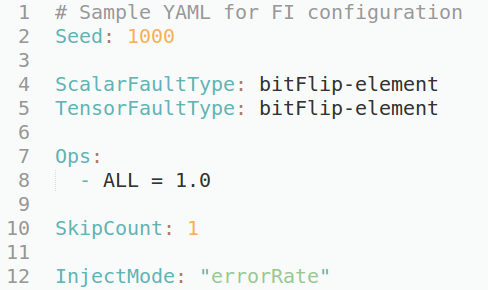}
    \caption{Example configuration file in YAML format}
    \label{fig:yaml}
\end{figure}

\sysname allows the user to configure it through a YAML interface. 
Figure~\ref{fig:yaml} shows a sample file for configuring \sysname in YAML format. This is loaded at program initialization, and is fixed for the entire fault injection campaign. The config file consists of the following fields:

\begin{itemize}[leftmargin=*]
    \item \textbf{Seed}: The random seed used in the fault injection experiments, for reproducibility purposes (this is optional).
    \item \textbf{ScalarFaultType}: The fault type to inject for scalar values (full list of types in Table~\ref{tbl:fault-types}).  We set this to {\em bitflip-Element}.
    \item \textbf{TensorFaultType}: The fault type to inject for tensor values (full list of types in Table~\ref{tbl:fault-types}). We set this to {\em bitflip-Element}.
    \item \textbf{InjectMode}: The mode of injection (list of modes in Table~\ref{tbl:fault-modes}). We set this to {\em errorRate}.
 \item \textbf{Ops}: This is a list of the \tfname operators that need to be injected, and the probability for injecting a fault into each operator when the mode is {\em errorRate}.  
 Probability values can range from 0 (never inject) to 1 (always inject). We choose {\em ALL}, which represents all operators. 
 \item \textbf{SkipCount}: This is an optional parameter for skipping the first `n' invocations of an operator before injection, where `n' can be any integer value, greater than 0.
\end{itemize}

\begin{table}
\small
\begin{center}
\begin{footnotesize}
\caption{List of fault types supported by \sysname}
\label{tbl:fault-types}
\begin{tabular}{|c|p{6cm}|}
\hline
Type & Explanation \\
\hline
None & Do not inject a fault \\
Zero & Change output of the target operator into all zeros \\
Rand & Shuffle \emph{all} data items in the output of the target operator into random values \\
Rand-element & Shuffle \emph{one} data item in the output of the target operator into a random value \\
bitFlip-element & Single bit-flip in \emph{one} data item in the output of the target operator \\
bitFlip-tensor & Single bit-flip in \emph{all} data items in the output of the target operator \\
\hline
\end{tabular}
\end{footnotesize}
\end{center}
\end{table}

\begin{table}
\caption{List of injection modes supported by \sysname}
\label{tbl:fault-modes}
\begin{footnotesize}
\begin{tabular}{|c|p{6cm}|}
\hline
\textbf{Mode} & \textbf{Meaning} \\
\hline
errorRate & Specify the error rate for different operator instances \\
\hline
dynamicInstance & Perform random injection on a randomly chosen instance of \emph{each} operation \\
\hline
oneFaultPerRun & Choose a single instance among all the operators at random  so that only one fault is injected in the entire execution \\
\hline
\end{tabular}
\end{footnotesize}
\end{table}

\section{Evaluation}
\label{sec:evaluation}
\label{sec:evaluation}

Our goal is to study how resilient are different ML applications (and datasets) to different fault configurations of \sysname, thereby demonstrating its utility. We first describe our experimental setup, followed by the research questions we ask in this study. We then present our results.

\subsection{Experimental Setup}


\textbf{ML applications:} We choose a total of 12 ML applications, 
12 of which are supervised learning applications 
(e.g., and deep neural networks like ResNet, VGGNet) that are commonly used in existing studies. In addition, we also choose an ML application used in the AV domain, i.e., comma.ai driving model. Table~\ref{tab:eval} lists the applications.

In addition to supervised models, \sysname can be used to inject faults into unsupervised models that use clustering, decision trees etc. We use one such application Generative Adversarial Networks (GAN) to show the effects of the injected faults visually. Because GANs do not have an expected output label, we do not consider it as part of the other experiments. 

\textbf{ML datasets:} We use 4 public ML datasets that are commonly used in ML studies. \emph{MNIST} dataset is a hand-written digits dataset (with 10 different digits). \emph{GTSRB} dataset is a dataset consisting of 43 different types of traffic signs. \emph{ImageNet} is a large image dataset with more than 14 million images in 1000 classes. In addition, we use a real-world driving dataset that is labeled with steering angles~\cite{drivingDataset}. 

For models that use the ImageNet dataset (ResNet-18 and SqueezeNet), we use the pre-trained model from since it is time-consuming to train the model from scratch. For the other models, we train them using the corresponding datasets. The datasets are summarized in Table~\ref{tab:eval}. The baseline accuracy of each model (without faults) is also provided for comparison. 


\textbf{Metrics:} 
We consider SDC rate as the metric for evaluating the resilience of ML applications. An SDC is a wrong output 
that deviates from the expected output of the program. 
SDC rate is the fraction of the injected faults that result in SDCs. 
For classifier applications, an SDC is any misclassification. 
However, the steering model {\em comma.ai} produces a continuous value as output. 
For this model, we use different threshold values for the deviations of steering angles to identify SDCs: 15, 30, 60 and 120 degrees~\cite{chen2019}. 
For the steering model, we use RMSE (root mean square error) and average deviation per frame as metrics to evaluate the model's accuracy - these are commonly used in ML studies int he AV domain~\cite{du2017self}.

\textbf{Experiments:}
For each benchmark, we perform $1000$ 
random FI experiments per fault configuration and input.
We choose 10 inputs for each application, and hence perform a total of $10,000$ fault injections per configuration and we use 14 different fault configurations.
We also calculate the error bars at the 95\% confidence interval for each experiment. 

\begin{figure}[t]
\centering
  \includegraphics[height=1.0in, width=3.5in]{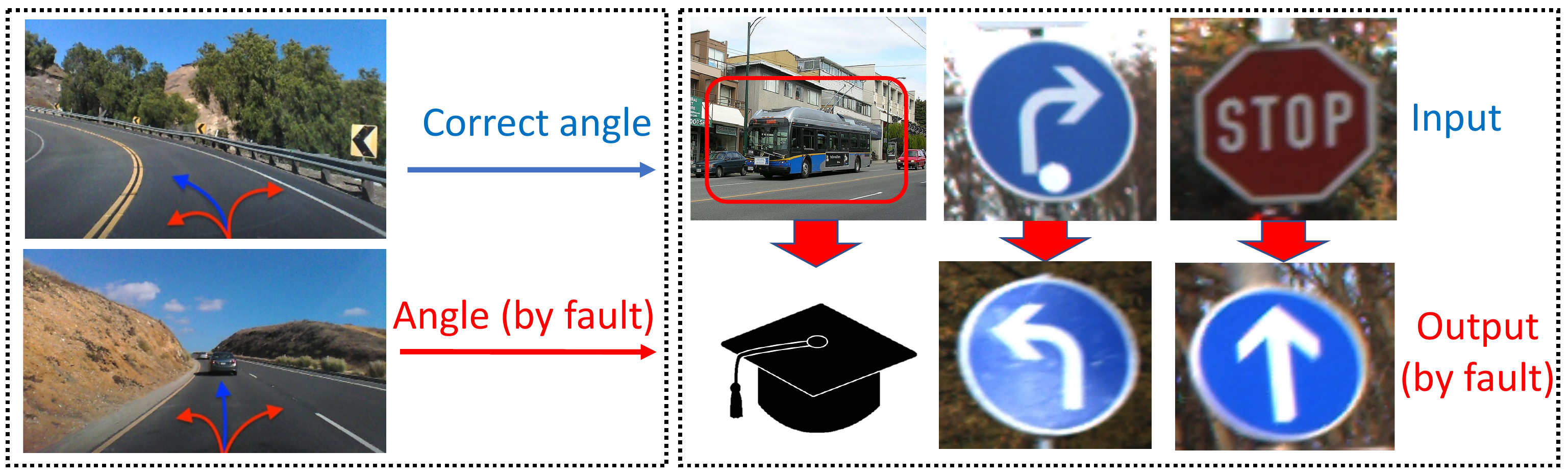}
\caption{Example of SDCs observed in different ML applications. Left box - Steering Model. Right box - Image Misclassifications.}
\label{fig:sdc-effect}
\end{figure}

%
Fig.~\ref{fig:sdc-effect} shows examples of some of the SDCs observed in our experiments for both the steering model and classification applications.
These may result in safety violations in AVs if they are not mitigated. With that said, we do not specifically distinguish safety-critical outcomes in SDCs. 

\begin{table}   
\small
\caption{ML applications and datasets used for evaluation. The baseline accuracy without faults is also provided.}
\label{tab:eval}
\centering
\begin{footnotesize}
\begin{tabular}{|c|c|c|c|}
\hline
{\bf ML model} & {\bf Dataset} & {\bf Dataset Description} & \textbf{Accuracy} \\
\hline
Neural Net & MNIST & Hand-written digits & 85.42\% \\
\hline
Fully Connected Net & MNIST & Hand-written digits & 97.54\% \\
\hline
LeNet & MNIST & Hand-written digits & 99\%\\
\hline
AlexNet & MNIST & Hand-written digits & 94\%\\
\hline 
CNN & MNIST & Hand-written digits & 95.74\% \\
\hline
Highway CNN & MNIST & Hand-written digits & 97.92\% \\
\hline
Recurrent NN & MNIST & Hand-written digits & 98.40\% \\
\hline 
VGG11 & GTSRB & Real-world traffic sign & 99.74\%\\ 
\hline 
ResNet-18  & ImageNet & General images & 62.66\% (top-1)\\
 & & & 84.61\% (top-5) \\
\hline   
SqueezeNet  & ImageNet & General images & 52.936\% (top-1)\\
 & & & 74.150\% (top-5)\\
\hline   
Comma.ai model \cite{commaai} & Driving & Real-world  & 24.12 (RMSE) \\
 & & driving frame & 12.64 (Avg. Dev.) \\

\hline
\end{tabular}
\end{footnotesize}
\end{table} 


\subsection{Research Questions}
We use different configurations of \sysname (shown in Table~\ref{tbl:fault-types} and Table~\ref{tbl:fault-modes}) for answering the following  Research Questions (RQs):

\textbf{RQ1:} What are the SDC rates of different applications under the {\em oneFaultPerRun} and {\em dynamicInstance} error modes? 

\textbf{RQ2:} For the {\em errorRate} mode, how do the SDC rates vary for different error rates? 

\textbf{RQ3:} How do the SDC rates vary for faults in different \tfname operations in the same ML application? 

\textbf{RQ4:} What is the performance overhead of \sysname? 


\subsection{Results}

We organize the results for the 11 ML models listed in Table \ref{tab:eval} by each RQ, and then show the results of the FI experiments for GANs. 
For RQ1 and RQ2, we choose \emph{all} the operations in the data-flow graph during the inference phase, which is a subset of operations in the \tfname graph. This is because many of the operations in the \tfname graph are for training, and are not executed during the inference phase (thus we do not inject faults into these operations). 
We also do not inject faults into those operations that are related to the input (e.g., reading the input, data preprocessing), as we assume that the inputs are correct as per our fault model.

\begin{figure}[t]
\centering
  \includegraphics[width=3.6in, , trim=4 4 4 4,clip]{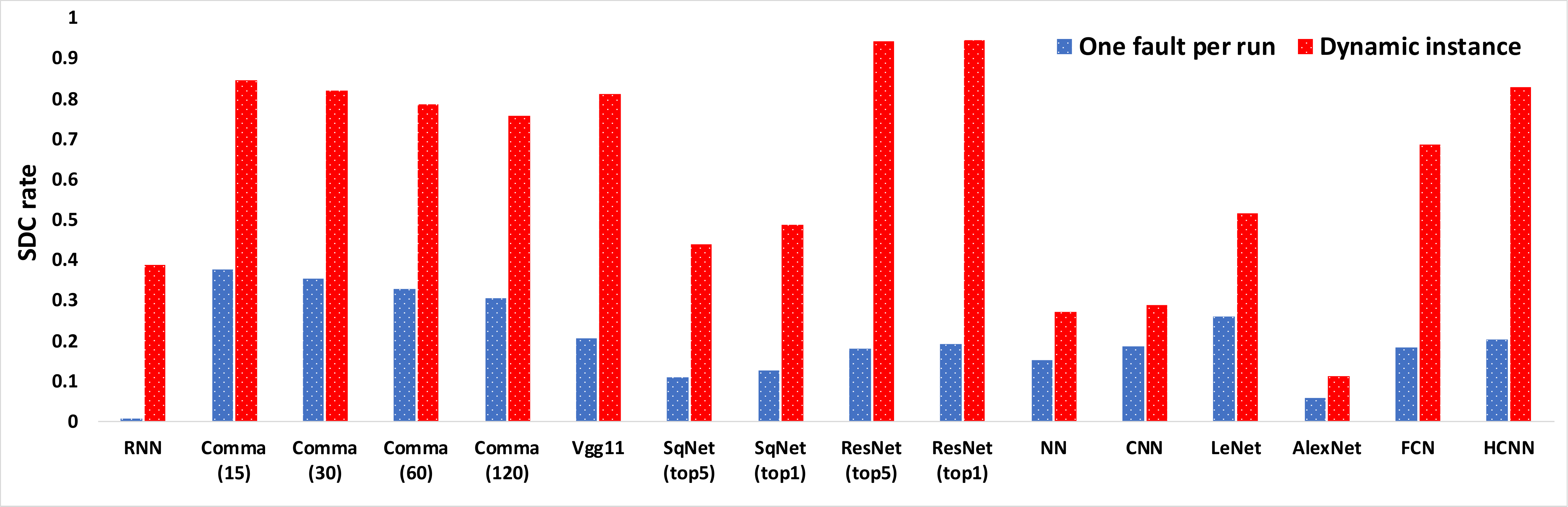}
\caption{SDC rates of different ML applications under bit-flip faults (\emph{from oneFaultPerRun and dynamicInstance injection modes}). Error bars range from $\pm 0.19\%$ to $\pm 2.45\%$ at the 95\% confidence interval}
\label{fig:bit-flip}
\end{figure}

\subsubsection{RQ1: Error resilience for different error modes} 
In this RQ, we study the effects of two different error modes, namely oneFaultPerRun and dynamicInstance. We choose single bit flip faults as the fault type for this experiment. 
Fig.~\ref{fig:bit-flip} show the SDC rates obtained across applications. 
We can see that different ML applications exhibit different SDC rates, and there is considerable variation.

We can also observe that there are differences between the two fault modes. 
For the dynamic instance injection mode, the SDC rates for all the applications are higher than those in the one fault per run mode. This is because in the dynamic instance mode, each type of operation will be injected at least once, while in the one fault per run mode, only one operator is injected in the entire application. Thus the applications present higher SDC rates for the former.

We also observe significant differences between applications within the one fault per run mode.
For example, the comma.ai driving model has a higher SDC rate than the classifier applications. This is because the output of the classifier applications are not dependent on the absolute values (instead classification probability is used).
Thus, the applications are still able to generate correct output despite the fault occurrence, and hence have higher resilience.
However, the comma.ai model predicts the steering angle, which is more sensitive to value deviations. For example, a deviation of 30 due to fault in the classification model will not cause an SDC as long as the predicted label is correct; whereas the deviation would constitute an SDC in the comma.ai model if we use the threshold=15 or 30.

In the one fault per run mode, we find that {\em RNN} exhibits the highest resilience (less than 1\% SDC rate).
This is because unlike feed-forward neural networks, RNN calculates the output not only using the input from the previous layer, but also the internal states from other cells. 
Under the single fault mode, the other internal states remain intact when the fault occurs at the output of the previous layer. 
Therefore, faults that occur in the feed-forward NNs are more likely to cause SDCs in this mode. 
However, under the dynamic instance injection mode, more than one fault will be injected.
As a result, some of the internal states are also corrupted, thus making the results prone to SDCs (e.g., RNN has around 38\% SDC rate).

We also find that {\em AlexNet} exhibits high resilience in the one-fault-per-run and dynamic instance injection modes. This is 
because AlexNet has many operations such as ADD, MUL, which are more resilient to faults (see Fig.~\ref{fig:sdc-per-operation}). 
Therefore, the proportion of operations that are more prone to SDCs (e.g., convolution operations, activation function) 
is not as high as that in other models such as VGG11.  


\begin{figure}
\centering
  \includegraphics[scale=0.31, trim=2 2 2 2,clip]{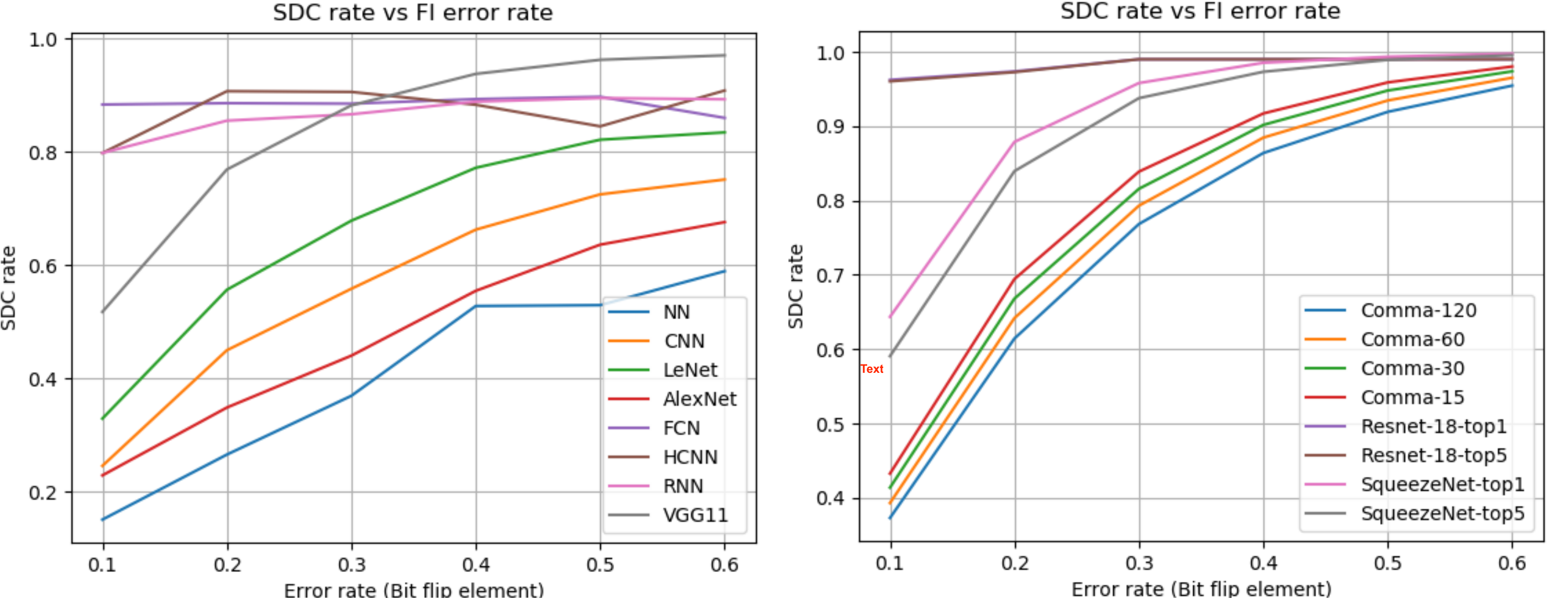}
\caption{SDC rates of different ML applications for various error rates \emph{(under bit flip element FI).}}
\label{fig:bit-flip-error-rate}
\end{figure}

\begin{figure}
\centering
  \includegraphics[scale=0.31, trim=2 2 2 2,clip]{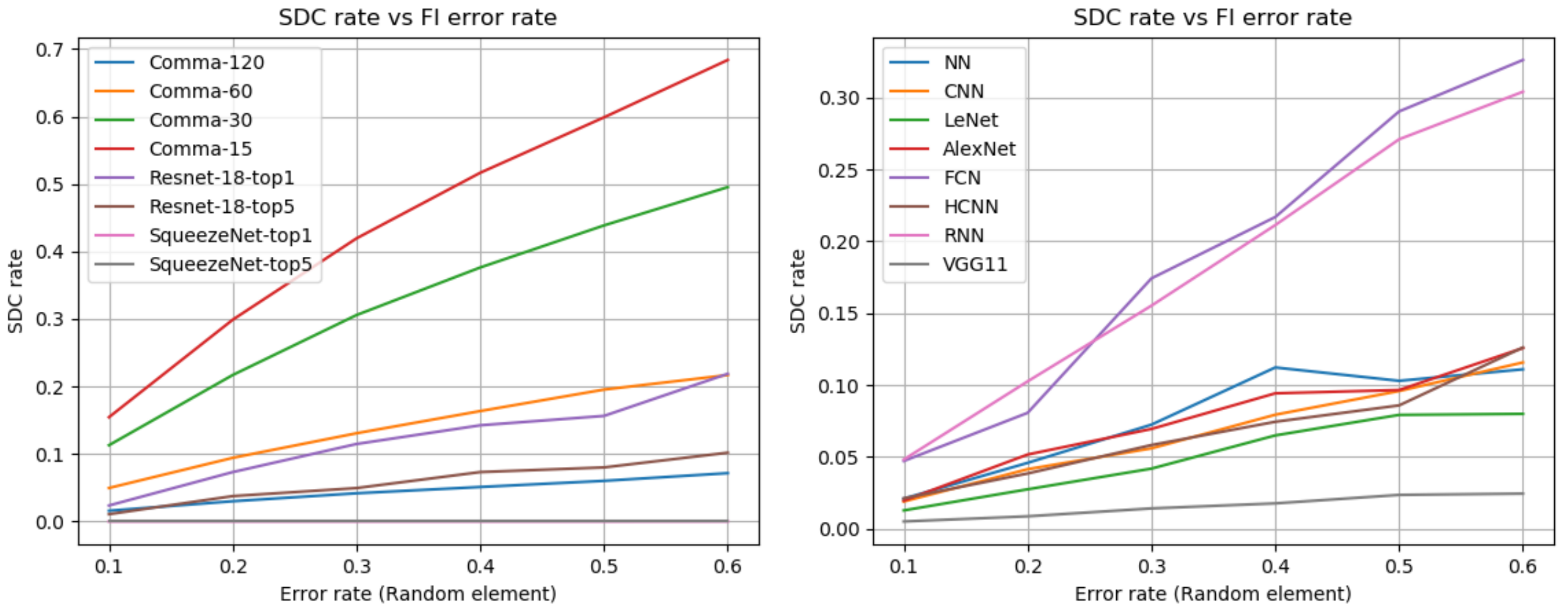}
\caption{SDC rates of different ML applications for various error rates \emph{(under random value replacement FI).}}
\label{fig:rand-error-rate}
\end{figure}

\subsubsection{RQ2: Error resilience under different error rates} 
In this RQ, we explore the resilience of different models for the {\em errorRate} injection mode. This mode allows us to vary the probability of error injection on a per-operator basis. We choose 2 different fault types for studying the effects of the error rate, namely bitFlip-element and Rand-element. 

Fig.~\ref{fig:bit-flip-error-rate} and Fig.~\ref{fig:rand-error-rate} show the variation SDC rates with error rates under both fault types.
As expected, we can observe that larger error rates results in higher SDC rates in all the applications, as more operations are injected.
However, compared with the results from the bit-flip FI, random value replacement results in lower SDC rates. 
This is likely because the random value causes lesser value deviation than the bit-flip fault type
 (in our implementation, we use the random number generator function from numpy library). 
Thus, a lower value deviation in this mode leads to lower SDC rates~\cite{li2017understanding,chen2019}.

Fig.~\ref{fig:bit-flip-error-rate} shows the variations of SDC rates of different ML applications with error rate under the bit-flip fault type. 
While we note that the SDC rates of all the applications grow along with the increase of error rates, 
we observe different applications have different rates of growth of SDCs.
In particular, we find that there are four outliers in the results for the bit-flip fault model 
(Fig.~\ref{fig:bit-flip-error-rate}), 
RNN, HCNN, ResNet and FCN, which exhibit significantly higher SDC rates than their counterparts. 
We find that the main reason is that these models have a significantly higher number of operations, 
thereby increasing the number of injected operators and resulting in higher SDC rates.

Likewise, in the case of the random replacement 
we find that the {\em SqueezeNet} applications
exhibit nearly flat growth in SDC rates with error rates,
 and that the SDC rates are consistently low. This is because 
faults need to cause large deviation in order to cause SDCs, which rarely occurs with the random replacement fault type.


\subsubsection{RQ3: SDC rates across different operations} 
\begin{figure}[t]
\centering
  \includegraphics[scale=0.35]{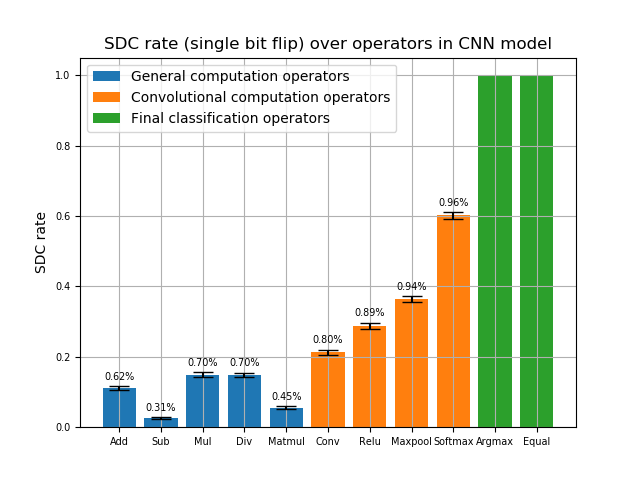}
\caption{SDC rates of different operations under bit-flip FI in the CNN model). Error bars range from $\pm 0.3077\%$ to $\pm 0.9592\%$ at 95\% confidence interval.}
\label{fig:sdc-per-operation}
\end{figure}

In this RQ, we study the SDC rates on different operations in the CNN model. The SDC rates are shown in Fig.~\ref{fig:sdc-per-operation}. 
It can be seen that faults in the convolution layer usually have higher SDC rates, compared with other operations (e.g., Sub). 

Moreover, we can see that operations such as SoftMax, ArgMax, Equal exhibit the highest SDC rates. 
In fact, the SDC rates on the ArgMax and Equal operations are nearly 100\%. This is because these operations are directly associated with the output, and thus faults in these operations are more likely to cause SDCs. On the other hand, operators such as {\em Sub}, {\em MatMul} have low SDC rates because faults in these operations 
are unlikely to propagate much.
For example, faults at the convolution layer are likely to propagate through the complex convolution operations, 
in which faults can quickly propagate and amplify.
However, faults at operations such as \emph{add} and \emph{multiply} might be masked \emph{before} propagating to the convolution layer; or occur \emph{after} the convolution layer. Therefore, due to the limited fault amplification effect, faults in these operations are less likely to cause SDCs.

\subsubsection{RQ4: Overhead.} In this question, 
we measure the execution time for the \tfname programs as a baseline for 50 predictions. 
Then we measure the time taken for 50 predictions after the \sysname instrumentation phase, but with fault injections disabled. 
These measurements are detailed in the \emph{Disable FI} column of the Table \ref{tab:runtime-1}. 
We then measure the time taken for 50 predictions with a single bit flip fault injected per run, and report this time in the \emph{Enable FI} column.
The first subcolumn `Inst.' in \emph{Overheads}, is the instrumentation overhead (difference between Disable FI and Baseline over the Baseline) and the second subcolumn `FI' gives the overhead incurred by fault injection alone (difference between Enable and Disable FI over Disable FI). 

As can be observed, the instrumentation overheads are relatively small, and range from 0.1x to 6.9x across applications. The fault injection overheads are much higher, ranging from 21x to 131x. This is because we are emulating the \tfname operators during fault injections in Python, and cannot benefit from the optimizations and low-level implementation of \tfname. However, the instrumentation phase itself incurs only modest overheads, with an average of 1.5x, when faults are not injected, in keeping with our minimal interference goal - this overhead is due to the runtime check for choosing which version of the operator to invoke (Section~\ref{sec:methodology}). 

While the overheads may seem high, we report the actual time taken by  the FI experiments to put this number in perspective.
In our experiments, the most time-consuming experiment is on the ResNet and Highway CNN models, which took less than 16 hours to complete. 
However, on average, most of our experiments took 3-4 hours to complete for injecting $10,000$ faults, which is reasonable. 


\begin{table}
\small
\caption{Overheads for the program \emph{(baseline)}; with instrumentation, without FI \emph{(disable FI)}; with FI \emph{(enable FI)}}
\label{tab:runtime-1}
\begin{center}
\begin{tabular}{|c|p{0.9cm}|p{0.9cm}|p{0.9cm}|p{1.8cm}|}
\hline
{\bf ML model} & {\bf Baseline} & {\bf Disable FI} & \bf{Enable FI} & {\bf Overheads} \\
  & (in s) &  (in s) & (in s) & \begin{tabular}{@{}c|c@{}}Inst. & FI\end{tabular} \\
\hline
NN & 0.06 & 0.16 & 13.50 & \begin{tabular}{@{}c|c@{}}1.6x & 83.34x\end{tabular}\\
\hline
FCN & 0.13 & 1.03 & 86.53 & \begin{tabular}{@{}c|c@{}}6.9x & 83x\end{tabular}  \\
\hline
LeNet & 0.105 & 0.22 & 17.44 & \begin{tabular}{@{}c|c@{}}1.1x & 78.27x\end{tabular}\\
\hline
AlexNet & 0.51 & 0.58 & 45.24 & \begin{tabular}{@{}c|c@{}}0.1x & 77x\end{tabular}\\
\hline
CNN & 0.23 & 0.23 & 25.94 & \begin{tabular}{@{}c|c@{}}0.1x & 107x\end{tabular}\\
\hline
HCNN & 0.44 &  1.02 & 134.56 & \begin{tabular}{@{}c|c@{}}1.3x & 131x\end{tabular}\\
\hline
RNN & 0.097 & 2.39 & 145 & \begin{tabular}{@{}c|c@{}}1.5x & 59.66x\end{tabular}\\
\hline 
VGG11 & 0.19 & 0.82 & 29.1 & \begin{tabular}{@{}c|c@{}}3.3x & 34.5x\end{tabular} \\ 
\hline   
SqueezeNet & 0.85 & 1 & 22 & \begin{tabular}{@{}c|c@{}}0.2x & 21x\end{tabular}\\
\hline   
ResNet & 2.72 & 3.76 & 300 & \begin{tabular}{@{}c|c@{}}0.4x & 78.78x\end{tabular}\\
\hline   
Comma.ai & 0.3 & 0.47 & 46 & \begin{tabular}{@{}c|c@{}}0.6x & 96.87x\end{tabular} \\
\hline   
\textbf{Average} & 0.59 & 1.06 & 78.66 & \begin{tabular}{@{}c|c@{}}1.5x & 77.3x\end{tabular}\\
\hline
\end{tabular}
\end{center}
\end{table}

\subsubsection{GAN FI results}
As mentioned, we show the GAN results separately as generated images do not have a labelled reference output, and we choose to not measure their SDC rates with another level of indirection from the discriminator.

The set of images in the top row, (ii) to (vi), are generated from setting the fault type to Rand-element. (ii) and (iii) are for one fault per run, and dynamic instance respectively. (iv) to (vi) are generated from the {\em errorRate} mode. (iv) is from setting the error rate to 25\%. We can see the fault progression clearly as the image becomes more difficult to decipher as the error rate increases. 
The second row  shows images obtained from similar configurations as the first row, with the only difference being that the fault type chosen is single bit flip. We observe that with bit flip in the operators, the resulting faults in images (vii) to (xi) tend to be more bipolar (i.e., have more black and white pixels than shades of grey). This is likely because with bit flips, the tensor values that store the image data are toggled between being present (1) at a pixel or being absent (0). As this error propagates into more operators, the computations performed amplify this effect and the resultant end images have strong activated regions of black or white. In the random replacement mode, the injected operations are replaced with random values and consequently the generated pixels to also exhibit any values within the range.


\begin{figure}[t]
\centering
  \includegraphics[scale=0.32]{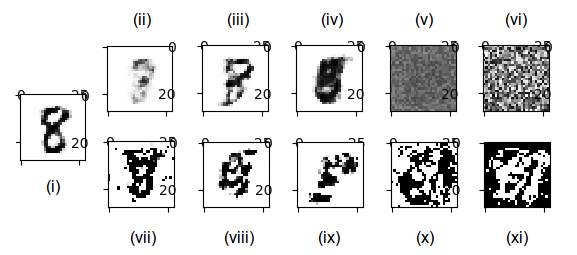}
\caption{Generated images of the digit 8 in the MNIST data-set under different configurations for GANs.
Top row represents the Rand-element model, while bottom row represents the single bit-flip model. Left center is with no faults.}
\label{fig:gan-8}
\end{figure}

\section{Conclusion} 
\label{sec:conclusion}
\label{sec:conclusion}

We present \sysname, a generic fault injection tool for ML applications written using the \tfname framework. \sysname is a configurable tool that can be easily integrated into existing ML applications. \sysname is portable, configurable, and fast. Further, it is 
also compatible with third party libraries that use \tfname. 
We use \sysname to study the resilience of 12 \tfname ML applications under different fault configurations, including one used in AVs. Our evaluation demonstrates the utility of \sysname in evaluating ML applications' resilience.





\bibliographystyle{IEEEtran}
\bibliography{IEEEabrv,lib}

\begin{thebibliography}{10}
\providecommand{\url}[1]{#1}
\csname url@samestyle\endcsname
\providecommand{\newblock}{\relax}
\providecommand{\bibinfo}[2]{#2}
\providecommand{\BIBentrySTDinterwordspacing}{\spaceskip=0pt\relax}
\providecommand{\BIBentryALTinterwordstretchfactor}{4}
\providecommand{\BIBentryALTinterwordspacing}{\spaceskip=\fontdimen2\font plus
\BIBentryALTinterwordstretchfactor\fontdimen3\font minus
  \fontdimen4\font\relax}
\providecommand{\BIBforeignlanguage}[2]{{%
\expandafter\ifx\csname l@#1\endcsname\relax
\typeout{** WARNING: IEEEtran.bst: No hyphenation pattern has been}%
\typeout{** loaded for the language `#1'. Using the pattern for}%
\typeout{** the default language instead.}%
\else
\language=\csname l@#1\endcsname
\fi
#2}}
\providecommand{\BIBdecl}{\relax}
\BIBdecl

\bibitem{li2018tensorfi}
G.~Li \emph{et~al.}, ``Tensorfi: A configurable fault injector for tensorflow
  applications,'' in \emph{IEEE International Symposium on Software Reliability
  Engineering Workshops (ISSREW)}.\hskip 1em plus 0.5em minus 0.4em\relax IEEE,
  2018.

\bibitem{banerjee2018hands}
S.~S. Banerjee \emph{et~al.}, ``Hands off the wheel in autonomous vehicles?: A
  systems perspective on over a million miles of field data,'' in \emph{48th
  Annual IEEE/IFIP International Conference on Dependable Systems and Networks
  (DSN)}.\hskip 1em plus 0.5em minus 0.4em\relax IEEE, 2018.

\bibitem{julian2016policy}
K.~D. Julian \emph{et~al.}, ``Policy compression for aircraft collision
  avoidance systems,'' in \emph{2016 IEEE/AIAA 35th Digital Avionics Systems
  Conference (DASC)}, 2016.

\bibitem{iso26262}
\BIBentryALTinterwordspacing
``Functional safety methodologies for automotive applications.'' [Online].
  Available:
  \url{https://www.cadence.com/content/dam/cadence-www/global/en_US/documents/solutions/automotive-functional-safety-wp.pdf}
\BIBentrySTDinterwordspacing

\bibitem{hsueh1997fault}
M.-C. Hsueh, T.~K. Tsai, and R.~K. Iyer, ``Fault injection techniques and
  tools,'' \emph{Computer}, vol.~30, no.~4, pp. 75--82, 1997.

\bibitem{stott2000nftape}
D.~T. Stott, B.~Floering, D.~Burke, Z.~Kalbarczpk, and R.~K. Iyer, ``Nftape: a
  framework for assessing dependability in distributed systems with lightweight
  fault injectors,'' in \emph{Proceedings IEEE International Computer
  Performance and Dependability Symposium. IPDS 2000}.\hskip 1em plus 0.5em
  minus 0.4em\relax IEEE, 2000, pp. 91--100.

\bibitem{carreira1998xception}
J.~Carreira, H.~Madeira, J.~G. Silva \emph{et~al.}, ``Xception: Software fault
  injection and monitoring in processor functional units,'' \emph{Dependable
  Computing and Fault Tolerant Systems}, vol.~10, pp. 245--266, 1998.

\bibitem{aidemark2001goofi}
J.~Aidemark, J.~Vinter, P.~Folkesson, and J.~Karlsson, ``Goofi: Generic
  object-oriented fault injection tool,'' in \emph{2001 International
  Conference on Dependable Systems and Networks}.\hskip 1em plus 0.5em minus
  0.4em\relax IEEE, 2001, pp. 83--88.

\bibitem{marinescu2009lfi}
P.~D. Marinescu and G.~Candea, ``Lfi: A practical and general library-level
  fault injector,'' in \emph{2009 IEEE/IFIP International Conference on
  Dependable Systems \& Networks}.\hskip 1em plus 0.5em minus 0.4em\relax IEEE,
  2009, pp. 379--388.

\bibitem{thomas2013llfi}
A.~Thomas and K.~Pattabiraman, ``Llfi: An intermediate code level fault
  injector for soft computing applications,'' in \emph{Workshop on Silicon
  Errors in Logic System Effects (SELSE)}, 2013.

\bibitem{wei2014quantifying}
J.~Wei, A.~Thomas, G.~Li, and K.~Pattabiraman, ``Quantifying the accuracy of
  high-level fault injection techniques for hardware faults,'' in \emph{2014
  44th Annual IEEE/IFIP International Conference on Dependable Systems and
  Networks}.\hskip 1em plus 0.5em minus 0.4em\relax IEEE, 2014, pp. 375--382.

\bibitem{abadi2016tensorflow}
M.~Abadi, P.~Barham, J.~Chen, Z.~Chen, A.~Davis, J.~Dean, M.~Devin,
  S.~Ghemawat, G.~Irving, M.~Isard \emph{et~al.}, ``Tensorflow: A system for
  large-scale machine learning,'' in \emph{12th $\{$USENIX$\}$ Symposium on
  Operating Systems Design and Implementation ($\{$OSDI$\}$ 16)}, 2016, pp.
  265--283.

\bibitem{paszke2019pytorch}
A.~Paszke, S.~Gross, F.~Massa, A.~Lerer, J.~Bradbury, G.~Chanan, T.~Killeen,
  Z.~Lin, N.~Gimelshein, L.~Antiga \emph{et~al.}, ``Pytorch: An imperative
  style, high-performance deep learning library,'' in \emph{Advances in Neural
  Information Processing Systems}, 2019, pp. 8024--8035.

\bibitem{keras}
\BIBentryALTinterwordspacing
``Keras.'' [Online]. Available: \url{https://keras.io/}
\BIBentrySTDinterwordspacing

\bibitem{kropp1998automated}
N.~P. Kropp, P.~J. Koopman, and D.~P. Siewiorek, ``Automated robustness testing
  of off-the-shelf software components,'' in \emph{Digest of Papers.
  Twenty-Eighth Annual International Symposium on Fault-Tolerant Computing
  (Cat. No. 98CB36224)}.\hskip 1em plus 0.5em minus 0.4em\relax IEEE, 1998, pp.
  230--239.

\bibitem{lanzaro2014empirical}
A.~Lanzaro, R.~Natella, S.~Winter, D.~Cotroneo, and N.~Suri, ``An empirical
  study of injected versus actual interface errors,'' in \emph{Proceedings of
  the 2014 International Symposium on Software Testing and Analysis}.\hskip 1em
  plus 0.5em minus 0.4em\relax ACM, 2014, pp. 397--408.

\bibitem{tf-popularity}
\BIBentryALTinterwordspacing
``Tensorflow popularity.'' [Online]. Available:
  \url{https://towardsdatascience.com/deep-learning-framework-power-scores-2018-23607ddf297a}
\BIBentrySTDinterwordspacing

\bibitem{alippi1995sensitivity}
C.~Alippi, V.~Piuri, and M.~Sami, ``Sensitivity to errors in artificial neural
  networks: A behavioral approach,'' \emph{IEEE Transactions on Circuits and
  Systems}, vol.~42, no.~6, 1995.

\bibitem{bettola1998high}
S.~Bettola and V.~Piuri, ``High performance fault-tolerant digital neural
  networks,'' \emph{IEEE transactions on computers}, no.~3, 1998.

\bibitem{li2017understanding}
G.~Li, S.~K.~S. Hari, M.~Sullivan, T.~Tsai, K.~Pattabiraman, J.~Emer, and S.~W.
  Keckler, ``Understanding error propagation in deep learning neural network
  (dnn) accelerators and applications,'' in \emph{Proceedings of the
  International Conference for High Performance Computing, Networking, Storage
  and Analysis}, 2017.

\bibitem{reagen2018ares}
B.~Reagen, U.~Gupta, L.~Pentecost, P.~Whatmough, S.~K. Lee, N.~Mulholland,
  D.~Brooks, and G.-Y. Wei, ``Ares: a framework for quantifying the resilience
  of deep neural networks,'' in \emph{55th Annual Design Automation
  Conference}, 2018.

\bibitem{chen2019}
Z.~Chen \emph{et~al.}, ``Binfi: An efficient fault injector for safety-critical
  machine learning systems,'' in \emph{Proceedings of the International
  Conference for High Performance Computing, Networking, Storage and
  Analysis}.\hskip 1em plus 0.5em minus 0.4em\relax ACM, 2019.

\bibitem{sabbagh2019evaluating}
M.~Sabbagh, C.~Gongye, Y.~Fei, and Y.~Wang, ``Evaluating fault resiliency of
  compressed deep neural networks,'' in \emph{2019 IEEE International
  Conference on Embedded Software and Systems (ICESS)}.\hskip 1em plus 0.5em
  minus 0.4em\relax IEEE, 2019, pp. 1--7.

\bibitem{ma2018deepmutation}
L.~Ma, F.~Zhang, J.~Sun, M.~Xue, B.~Li, F.~Juefei-Xu, C.~Xie, L.~Li, Y.~Liu,
  J.~Zhao \emph{et~al.}, ``Deepmutation: Mutation testing of deep learning
  systems,'' in \emph{2018 IEEE 29th International Symposium on Software
  Reliability Engineering (ISSRE)}.\hskip 1em plus 0.5em minus 0.4em\relax
  IEEE, 2018, pp. 100--111.

\bibitem{iyer1993experimental}
R.~Iyer, D.~Tang \emph{et~al.}, ``Experimental analysis of computer system
  dependability,'' 1993.

\bibitem{drivingDataset}
\BIBentryALTinterwordspacing
``Driving dataset.'' [Online]. Available:
  \url{https://github.com/SullyChen/driving-datasets}
\BIBentrySTDinterwordspacing

\bibitem{du2017self}
S.~Du,  \emph{et~al.}, ``Self-driving car steering angle prediction based on
  image recognition,'' \emph{Department of Computer Science, Stanford
  University, Tech. Rep. CS231-626}, 2017.

\bibitem{commaai}
\BIBentryALTinterwordspacing
``comma.ai's steering model.'' [Online]. Available:
  \url{https://github.com/commaai/research}
\BIBentrySTDinterwordspacing

\end{thebibliography}

\end{document}